\documentclass[sort&compress,final]{aipproc}
\usepackage{natbib}
\layoutstyle{6x9}

\begin{document}

\title{Jet Quenching and Holographic Thermalization}

\author{E. Caceres$^{1,2}$, A. Kundu$^2$, B. M\"uller$^3$, D. Vaman$^4$, D.-L. Yang$^3$}{
  address={$^1$Facultad de Ciencias, Universidad de Colima, Bernal Diaz del Castillo 340, Colima, Mexico. \\
$^2$Theory Group, Department of Physics,
University of Texas at Austin, Austin, TX 78712, USA.\\ 
$^3$Department of Physics, Duke University, Durham, North Carolina 27708, USA\\
$^4$Department of Physics, University of Virginia, Charlottesville, Virginia 22904, USA}
}

\classification{12.38Mh, 12.90.+b, 04.70.-s}

\keywords{Quark-Gluon Plasma, AdS/CFT correspondence, Gravitational Collapse}

\begin{abstract}
We employ the AdS/CFT correspondence to investigate the thermalization of the strongly-coupled plasma and the jet quenching of a hard probe traversing such a thermalizing medium. 
\end{abstract}

\maketitle

The AdS/CFT correspondence \cite{Maldacena:1997re,Witten:1998qj,Witten:1998zw,Gubser:1998bc,Aharony:1999ti}, a duality between the strongly coupled $\mathcal{N}=4$ super Yang-Mills theory and type IIB supergravity, has been widely used to study the qualitative features of the quark gluon plasma (QGP) generated from the relativistic heavy ion collisions. In the gravity dual, the thermalization of the medium corresponds to the gravitational collapse and the formation of a black hole, which has been currently investigated in \cite{Lin:2008rw,Chesler:2008hg,Bhattacharyya:2009uu,Balasubramanian:2010ce,Balasubramanian:2011ur,Garfinkle:2011hm,Garfinkle:2011tc}.
A light probe traveling in the strongly coupled medium could finally dissipate, which leads to a maximum stopping distance \cite{Chesler:2008uy,Gubser:2008as,Hatta:2008tx,Arnold:2011qi}. The stopping distance could qualitatively characterize the jet quenching of light probes.    

In this note, we briefly present our work and highlight the results in \cite{yang,yang2}. We utilize the AdS-Vaidya spacetime proposed in \cite{Balasubramanian:2010ce,Balasubramanian:2011ur}, which describes a falling mass shell, to analyze the isotropic thermalization. The metric is induced by the leading order perturbation of a weak dilaton field\cite{Bhattacharyya:2009uu}.
In Eddington-Finkelstein (EF) coordinates, the metric is given by 
\begin{eqnarray}\label{efvaidya}
ds^2=\frac{1}{z^2}\left(-(1-m(v)z^4)dv^2-2dvdz+dx_i^2\right),
\end{eqnarray} 
where $m(v)=\frac{M}{2}(1+\tanh(v/v_0))$ represents the mass function of the shell and $v$ denotes the EF time coordinate.
In the thin-shell limit (ie. $v_0\rightarrow 0$), the AdS-Vaidya spacetime will be separated into two regimes. The exterior of the shell is dictated by the AdS-Schwarzschild metric, while the interior is governed by the quasi-AdS geometry, which corresponds to the vacuum with a nonzero gravitational potential \cite{yang}. In Poincare coordinates, the metric is written as
\begin{eqnarray}\label{metric}
ds^2=\left\{
\begin{array}{ll}
\frac{1}{z^2}\left(-F(z)dt^2+\frac{dz^2}{F(z)}+dx_i^2\right) & \mbox{if } z < z_0(v>0), \\
\\
\frac{1}{z^2}\left(-F(z_0)^2dt^2+dz^2+dx_i^2\right) & \mbox{if }z>z_0(v<0),
\end{array} \right.
\end{eqnarray}
where $z_0$ denotes the position of the shell and $F(z)=1-z^4/z_h^4$ for $z_h=M^{-1/4}=(\pi T)^{-1/4}$ being the future horizon. The position of the shell is determined by the redshift factor $F(z_0)$. By tracking the position of the shell, a rapid thermailization time scaled by the temperature is found, $\tau T\approx 0.55$, which is consistent with the approximated thermalization time scale at the RHIC energy \cite{yang}.    

When the medium starts to thermalize, an energetic light probe could be created simultaneously. In the gravity dual, the light probe is characterized by a massless particle falling along the null geodesic from the boundary to the bulk by using the WKB approximation, which eventually falls into the horizon and results in the thermalization of an image current on the boundary \cite{Hatta:2008tx,Arnold:2011qi}. Given that the shell is infinitesimally thin, the particle travels most of time in the exterior of the shell; thus the stopping distance in this case is the same as that found in the thermalized medium. On the contrary, when the shell has finite thickness, the light probe could travel further within the shell. However, by solving the Einstein equations in EF coordinates numerically, it turns out that the stopping distance of the light probe is approximately equal to that in the thermalized medium. In general, the jet quenching of an energetic light probe may not be affected by the thermalization process of the plasma \cite{yang}.

\begin{figure}[h]
\begin{minipage}{7.5cm}
\begin{center}
{\includegraphics[width=7.5cm,height=4cm,clip]{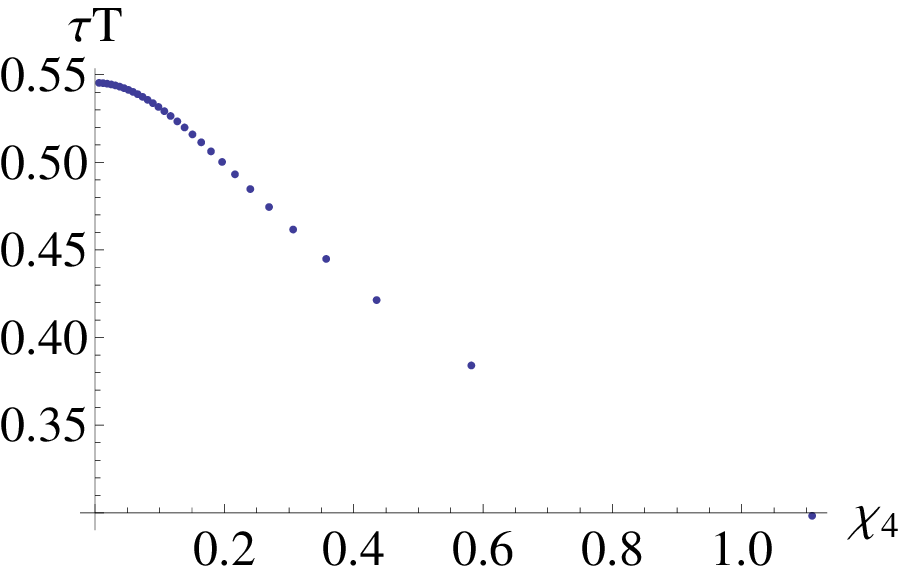}}
\caption{}
\end{center}
\end{minipage}
\begin{minipage}{7.5cm}
\begin{center}
{\includegraphics[width=7.5cm,height=4cm,clip]{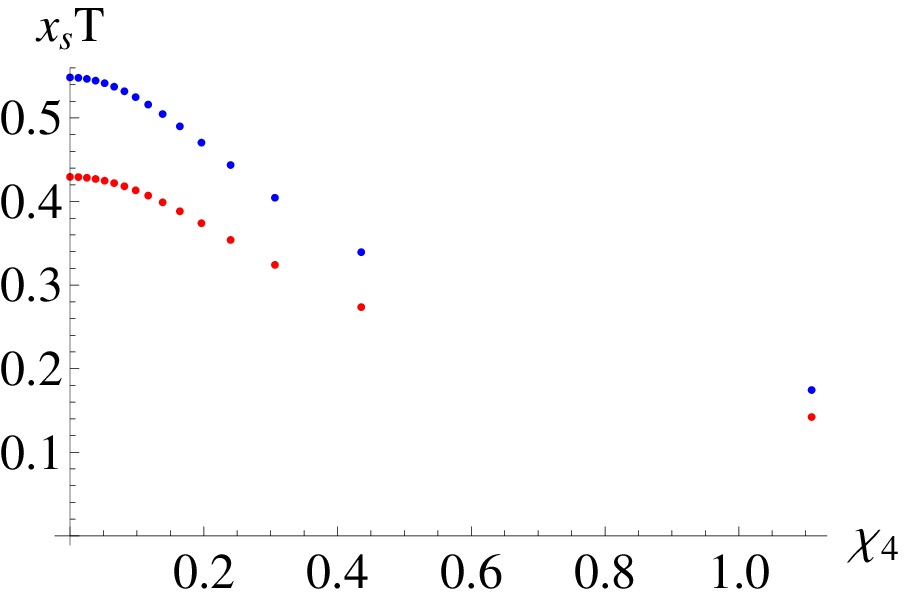}}
\caption{The left panel shows the thermalization time $\tau$ with different values of chemical potential \cite{yang2}. On the right panel, the red and blue points represent the stopping distances with different values of chemical potential in AdS-RN and AdS-RN-Vaidya spacetimes, respectively \cite{yang2}.}\label{xsmu5d}
\end{center}
\end{minipage}
\end{figure}

Furthermore, the generalization of the AdS-Vaidya spacetime by introducing a charged shell is presented in \cite{Caceres:2012em}; the generalized metric, AdS-Reissner-Nordstr\"om-Vaidya (AdS-RN-Vaidya) geometry corresponds to a thermalizing plasma with a nonzero chemical potential. By following the same approach in \cite{yang} to track the falling shell in Poincare coordinates, the thermalization time of the medium with a nonzero chemical potential can be found \cite{yang2}. As shown in Fig.\ref{xsmu5d}, the thermalization time decreases when the chemical potential is increased, where $\chi_4=(4\pi T)^{-1}\mu$ for $\mu$ being the chemical potential. 

In addition, the jet quenching of a light probe such as a virtual gluon traveling in such a non-equilibrium plasma can be studied by applying the approach of a falling string introduced in \cite{Gubser:2008as}. The tip of the string is represented by a massless particle moving along the null geodesic, which leads to a maximum stopping distance similar to the approach of taking the WKB approximation to the falling wave packet induced by an image jet on the boundary \cite{Hatta:2008tx,Arnold:2011qi}. Nevertheless, the energy of the gluon is now encoded in the initial profile of the string, where $E\sim 1/z_I$ for $z_I$ being the initial position of the tip of the string. When $z_I>0$, the gluon will carry finite energy comparable to the thermalization temperature. Such gluons serve as soft probes and their stopping distances could be influenced by the thermalization process. As shown in Fig.\ref{xsmu5d}, the stopping distances of the gluons as well decrease when the chemical potentials are increased in both thermalizing and thermalized media. Also, for the soft gluon, its stopping distance in the thermalizing medium is larger than that in the thermalized case, which is expected since the probe in the former scenario travels for a longer time in the vacuum. 

In conclusion, by using the gauge/gravity duality in the framework of the AdS-Vaidya type geometries, we found that the medium thermalizes rapidly, which characterizes the qualitative feature of a strongly coupled plasma. Moreover, the jet quenching of a light probe with infinite energy compared to the thermalization temperature is insusceptible to the thermalization process of the medium, while the jet quenching of the soft probe in the thermalizing plasma is reduced. When the chemical potential of the medium is increased, the thermalization of both the medium and the probe is accelerated, which could come from the enhanced scattering led by the increase of the density of the plasma \cite{yang2}.              
\begin{theacknowledgments}
The authors thank P. Arnold for useful discussions. We also thank the organizers 
of the Eleventh Conference on the Intersections of Particle and Nuclear Physics 
(CIPANP 2012).
\end{theacknowledgments}

\bibliographystyle{aipproc}   

\end{document}